\documentclass[iop, revtex4]{emulateapj}
\usepackage{natbib}
\usepackage{array}

\usepackage{graphicx}
\usepackage{latexsym}
\usepackage{amsfonts,amsmath,amssymb}
\usepackage{url}
\usepackage[utf8]{inputenc}

\usepackage{hyperref}
\usepackage{placeins}
\hypersetup{colorlinks=false,pdfborder={0 0 0},}

\shortauthors{Adam Greenberg and Jean-Luc Margot}

\begin{document}

\title{Improved Algorithms for Radar-Based Reconstruction of Asteroid Shapes}

\shorttitle{Improved Algorithms for Radar-Based Reconstruction of Asteroid Shapes}

\author{Adam H. Greenberg}
\author{Jean-Luc Margot}

\affil{University California, Los Angeles}

\begin{abstract}

We describe our implementation of a global-parameter optimizer and
Square Root Information Filter (SRIF) into the asteroid-modelling
software {\tt shape}.  We compare the performance of our new optimizer
with that of the existing sequential optimizer when operating on
various forms of simulated data and actual asteroid radar data.  In
all cases, the new implementation performs substantially better than
its predecessor: 
it converges faster, produces shape models that are
more accurate, and solves for spin axis orientations more reliably.  We
discuss potential future changes to improve {\tt shape}'s fitting
speed and accuracy.

\end{abstract}


\bibliographystyle{apj}

\section{Introduction}
\label{sec-intro}
Earth-based radar is a powerful tool for gathering information about
bodies in the Solar System. Radar observations 
can dramatically improve the determination of 
the physical properties and orbital elements of
small bodies (such as asteroids and comets).  
An important development in the past two decades has been the
formulation and implementation of algorithms for asteroid shape
reconstruction based on radar
data~\citep{huds93,hudson1994,ostr95,hudson1995}.  
This problem is not trivial because it requires the joint estimation
of the spin state and shape of the asteroid.  Because of the nature of
radar data, recovery of the spin state depends on knowledge of the
shape and vice versa.  Even with perfect spin state information,
certain peculiarities of radar images (such as the two-to-one or
several-to-one mapping between surface elements on the object and
pixels within the radar image) make recovery of the physical shape
challenging~\citep{ostro1993b}.  This is a computationally intensive
problem, potentially involving hundreds to thousands of free
parameters and millions of data points.

Despite the computational cost, astronomers are keen on deriving shape
and spin information from asteroid radar images.  The most compelling reason to
do so is the fact that radar is the only Earth-based technique that
can produce detailed three-dimensional information of near-Earth objects.  This is
possible because radar instruments achieve spatial resolutions that
dramatically surpass the diffraction limit.  In other words, radar
instruments can resolve objects substantially smaller than the
beamwidth of the antenna used to obtain the images. For example,
the Arecibo telescope, the primary instrument used for the data
presented in this paper, has a beamwidth of ${\sim}2$ arcminutes
at the nominal 2380 MHz frequency of the radar. Yet 
observers can easily gather shape information to an accuracy
of decameters for objects several millions of kilometers from Earth,
achieving an effective spatial resolution of ${\sim}1$ milliarcsecond.

Radar has other advantages as well. Unlike most observational
techniques inside the Solar System, radar does not rely on any
external sources of light, be it reflected sunlight, transmitted
starlight, or thermal emission. This human-controlled illumination
allows for greater flexibility with respect to the observations. In addition,
because of the wavelengths involved, radar observations can be performed
during the day, further enhancing this flexibility.
Radar also has the ability to probe an object's sub-surface
properties, which can give important information about the object such
as porosity, surface and sub-surface dielectric constant, and the
presence of near-surface ice.

\begin{figure*}[t!]
\begin{center}
\includegraphics[width=1.8\columnwidth]{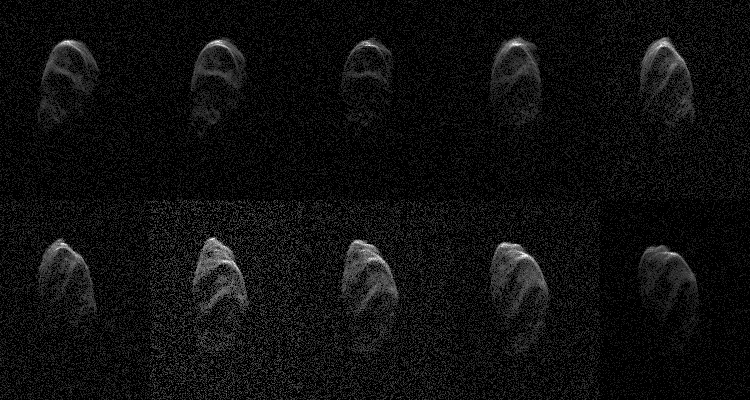}
\caption{A time-series of range-Doppler images of the asteroid
  2000~ET70~\citep{naidu2013}, starting in the top left and proceeding
  to the right.  The epochs of consecutive images are separated by 18
  minutes. Distance from the observer increases downwards, and Doppler
  increases to the right.}
\end{center}
\label{fig-radar}
\end{figure*}

Asteroid shape data are important for various reasons.  For certain
asteroids, reliable determination of an orbital future cannot be
determined without shape 
and spin information. The Yarkovsky effect, for example, can change an
asteroid's semi-major axis 
at a rate of 
${\sim}10^{-4}$ AU/My for km-sized
objects~\citep{vokrouhlicky2000,bott06areps,nuge12yark}.  
This effect occurs because the rotating body absorbs sunlight and then
re-emits that light in a non-sunward direction, resulting in a gentle
perturbation to the asteroid's orbit.  The Yarkovsky effect is greatly
dependent on the shape of the object, since re-emission of absorbed
sunlight is a surface phenomenon. It is responsible for the largest source of uncertainty in trajectory
predictions for near-Earth asteroids (NEAs) with sizes under 2 km, and
it must be taken into account when evaluating impact
probabilities~\citep{giorgini2002,ches14,farnocchia2013}.

Knowledge of the shape also provides clues about the formation and
interaction history of asteroids.  For example, radar-derived shapes
of asteroids have been instrumental in identifying binary asteroids
and contact binaries, which represent ${\sim}16\%$ and ${\sim}10\%$ of the
population, respectively~\citep{marg02s,benner2008}.  They have also
provided strong evidence that NEA binaries form by a spin-up and mass
shedding process~\citep{marg02s,ostr06}.  
For single asteroids, knowledge of the morphology guides
interpretation of the collisional history and surface modification
processes.

Shapes also affect spin evolution during two-body interactions (e.g.,
torques during close planetary encounters) and orbital evolution of
binary NEAs (e.g., tidal, gravitational, and non-gravitational
interactions between
components)~\citep[e.g.,][]{marg02s,ostr06,sche06,cuk10,jaco14,naid15}.
Finally, shapes are needed when calculating the gravity environment
near asteroids, which is of special importance for proximity
operations~\citep{fujiwara2006,naidu2013,nola13,taka14}.

The determination of an asteroid's spin state from radar data is
equally valuable.  In contrast to lightcurve period determinations,
which are neither sidereal nor synodic, the radar-based measurements
yield sidereal periods.  These estimates are needed to test the
agreement between physical theories and observations, e.g., the change
in asteroid spin rate due to sunlight~\citep[e.g.,][]{tayl07,lowr07}
and subsequent shape evolution~\citep[e.g.,][]{harr09,fahn09}.  Proper
modeling of the Yarkovsky perturbations to an asteroid's heliocentric
orbit or to the evolution of binary orbits~\citep[e.g.,][]{marg15AIV}
also require knowledge of the spin state.  Finally, important insights
can be gained about asteroid physical properties and collisional
evolution from the spin distributions of both regular rotators and
non-principal-axis rotators~\citep[e.g.,][]{prav02AIII}.

\section{Current Method}
\label{sec-current}
Asteroid shapes and spin states are currently modeled using the {\tt
shape} software package \citep{huds93,magr07}. {\tt shape} takes a
model for the asteroid, which is based on both shape and spin
parameters, as well as scattering behavior, and projects that model
into the same space as that of the radar observables. This space,
called the range-Doppler space, has dimensions of range and
line-of-sight velocity (figure~\ref{fig-radar}).  {\tt shape} can also
handle optical lightcurve data when fitting for asteroid shapes, but
we did not use this capability in this paper.

{\tt shape} then compares the mapping of the model into this space to
the radar observables, and makes changes to the model parameters in an
attempt to minimize the sum of squares of residuals.  {\tt shape} uses
increasing model complexity to build up a representation for the
asteroid, from a basic ellipsoid model to capture gross features, to a
spherical harmonic model which can represent finer surface elements
(See section~\ref{sec-fake}) and finally a model based on contiguous
triangular facets (hereafter vertex model).  The spin state is
generally estimated in the early stages of the shape fitting -- this
is normally done by using trial values of the spin state while
simultaneously fitting for the shape itself.

{\tt shape} currently uses a Sequential Parameter Fit (SPF) mechanism
to adjust the model following a comparison between the model
projection and the radar observables.  SPF minimizes $\chi^2$ using a
``bracket and Brent" method \citep{numericalrecipes} -- for each
iteration, this process minimizes $\chi^2$ for variations in that
individual parameter only, while all other parameters are held
constant.  This process is not only slow, but it also does not
guarantee convergence on a global minimum, or even the nearest local
minimum, because minimization always progresses along a single
parameter axis at a time. We have worked towards replacing the SPF
currently implemented in {\tt shape} with a modified Square Root
Information Filter (SRIF), as outlined in section~\ref{sec-srif}.

\section{Solution via Normal Equations}
\label{sec-nem}

Before detailing the mechanics of the SRIF, it is worth discussing the
Normal Equations Method (NEM), to which SRIF is related ~\citep{numericalrecipes}.  A classical NEM
minimizes the weighted residuals between a model and data with noise
assumed to be Gaussian by determining the direction in parameter space
in which $\chi^2$ is decreasing fastest. Specifically, suppose one has
a set of $m$ observables, $\vec{z}$, with weights that are the
diagonal elements of an $m \times m$ matrix $W$, and a model function
$\vec{f}(\vec{x})$, where $\vec{x}$ is an $n$-dimensional parameter
vector. Assuming independent data points with Gaussian-distributed
errors, the probability of the model matching the data is given by

\[p(\vec{f}(\vec{x})\ |\ \vec{z}) \propto p(\vec{z}\
|\ \vec{f}(\vec{x})) \propto \exp( -\frac{1}{2}\vec{R}^\intercal W
\vec{R}),\] 
where $\vec{R} = \vec{z} - \vec{f}(\vec{x})$. Therefore maximizing the
model probability is the same as minimizing the value
\[\chi^2(\vec{x}) = \vec{R}^\intercal W \vec{R}.\] 
Perturbing $\vec{x}$ by some amount, $\vec{\delta x}$, and minimizing
$\chi^2(\vec{x})$ over $\vec{\delta x}$ yields
\[(A^\intercal W A)\vec{\delta x} = A^\intercal W \vec{R},\]
where \[A = \frac{\partial \vec{R}}{\partial
\vec{x}}.\] 
Thus, changing one's parameter vector by
\begin{equation}
\label{eq-normal}
 \vec{\delta x} = (A^\intercal W A)^{-1} A^\intercal W \vec{R}
 \end{equation}
yields a decrease in $\chi^2(\vec{x})$. For non-linear systems, this
process is repeated multiple times until the change in $\chi^2$ from
one iteration to the next has passed below a fiducial fraction. Equation~\ref{eq-normal}
is also known as the weighted normal equation.

A major issue with NEM is the computation of the inverse of the matrix
$A^\intercal W A$. This matrix has $n^2$ elements and thus can be
quite large for a model with many parameters. In addition, numerical
stability can be a serious issue -- $A^\intercal W A$ may be
ill-conditioned, and thus taking the inverse can result in numerical
errors (see Appendix).

One way to quantify the issue of numerical stability is by using the condition number
$\kappa(M) $, where \linebreak
$\kappa(M) \equiv ||M|| * ||M^{-1}||$.  A smaller $\kappa(M)$ 
corresponds to a better conditioned matrix $M$,
meaning that fewer errors will accrue in the calculation of $M^{-1}$.

Since \[\kappa(A^\intercal W A) \propto \kappa(A)^2 \] and \[\kappa(A) > 1 \] for 
non-orthogonal matrix A, the classical NEM increases the risk of
numerical instabilities

Finally, for problems involving a very large number of observations
and model parameters, even the calculation of $(A^\intercal W A)$ is
non-trivial, as this matrix multiplication scales like $m^2n$. The
number of observations needed for an asteroid shape reconstruction
typically number in the millions, with potentially $10^2 - 10^5$ free
parameters.

\subsection{Square Root Information Filter}
\label{sec-srif}
The Square Root Information Filter (SRIF) was originally developed by
Bierman in 1977 (\cite{bierman1977}, \cite{lawson1995}). The algorithm
minimizes $\chi^2$ for time series data with Gaussian errors, and is
inspired by the Kalman filter algorithm. SRIF is more stable, more
accurate, and faster than the algorithm currently used in
{\tt shape}. SRIF is also more numerically stable (and, in some
cases faster) than a solution via normal equations. Our implementation
of SRIF includes some changes to the original algorithm, which will be
discussed in section~\ref{sec-optimizations}.

SRIF gets around all the problems described above by utilizing matrix
square roots and Householder operations (see \cite{bierman1977},
pg. 59) to increase the numerical stability when determining
$\delta\vec{x}$. Instead of minimizing $\chi^2$, SRIF minimizes
\[Q = (\chi^2)^{\frac{1}{2}} = ||W^{\frac{1}{2}}\vec{R}||,\]
where $W^{\frac{1}{2}}$ is the square root of the matrix $W$, defined
such that \[W = W^{\frac{1}{2}}W^{\frac{1}{2}}.\] 
In general, the square root of a matrix is multivalued, however since
$W$ is positive-semidefinite, all square roots are real.  We select
the positive root by convention.

Then, along similar lines as NEM, a change of $\vec{\delta x}$ is
introduced to the parameter vector $\vec{x}$, and $Q' =
Q(\vec{x}+\vec{\delta x})$ is minimized over this change.

$Q'$ is smallest when \[||W^{\frac{1}{2}} \vec{R}(\vec{x} +
\vec{\delta x})|| \approx ||W^{\frac{1}{2}}(\vec{R}(\vec{x}) +
A\vec{\delta x})|| \] \[= ||W^{\frac{1}{2}}\vec{R}(\vec{x}) +
W^{\frac{1}{2}} A\vec{\delta x}||\] is minimized.

A matrix $H$ is defined such that $HW^{\frac{1}{2}} A$ is upper
triangular. $H$ is orthogonal and can be generated by the product of
$n$ Householder operation matrices.  Note that the orthogonality of
$H$ guarantees that
\[ ||W^{\frac{1}{2}}\vec{R}(\vec{x}) + W^{\frac{1}{2}} A\vec{\delta
x}|| = ||H(W^{\frac{1}{2}}\vec{R}(\vec{x}) + W^{\frac{1}{2}}
A\vec{\delta x})||\] \[= || HW^{\frac{1}{2}}\vec{R}(\vec{x}) +
HW^{\frac{1}{2}} A\vec{\delta x} ||.\]  
From the definition of $H$, $HW^{\frac{1}{2}} A$ can be rewritten as
\[ HW^{\frac{1}{2}} A = \left( \begin{array}{c}
A' \\ Z \end{array} \right),\] 
where $A'$ is an $n \times n$ upper-triangular matrix, and $Z$ is the
$(m-n) \times n$ zero-matrix.  Then, rewriting
\[HW^{\frac{1}{2}}\vec{R}(\vec{x}) = \left( \begin{array}{c}
\vec{R_x'} \\ \vec{R_z'} \end{array} \right),\] 
where $R_x'$ and $R_z'$ are $m \times 1$ and $(n-m) \times 1$ arrays,
respectively, yields
\[ Q' = \left|\left|\left( \begin{array}{c}
\vec{R_x'} + A'\vec{\delta x} \\
\vec{R_z'}  + Z\vec{\delta x} \end{array} \right)\right|\right| 
= \left|\left|\left( \begin{array}{c}
\vec{R_x'} + A'\vec{\delta x} \\
\vec{R_z'} \end{array} \right)\right|\right|
.\] 
This is clearly minimized over $\delta x$ when 
\[ \vec{R_x'} =
-A'\vec{\delta x} \] 
or 
\begin{equation}
\vec{\delta x} = -A'^{-1} \vec{R_x'}.
\end{equation}

Since $A'$ is upper triangular, its inverse can be easily calculated,
and singularity can be trivially detected.  Furthermore, the condition
number of the inverted matrix is proportional to $\kappa(A)$, as
opposed to $\kappa(A)^2$ in the NEM case.

Finally, note that $A^\intercal W A$ is never calculated, which, as
mentioned in section~\ref{sec-nem}, is a computationally intensive
calculation.

\section{Additions to SRIF}
\label{sec-optimizations}
\subsection{Optimizations}
The number of operations necessary to generate the Householder matrix
$H$ grows as O($n^2(m-n)$), where the number of observations $m$
always exceeds the number of parameters $n$.  Although this growth
profile is favorable with respect to $m$ when compared to that of NEM
(O($m^2n$)), it becomes problematic for high resolution models (large
$n$).  To maintain good performance in large $n$ situations, we have
implemented three main optimizations to the standard SRIF.

Our first addition is to run the matrix triangularization
simultaneously on multiple cores, which results in a significant
speed-up.  Note that although Householder matrices are generated
iteratively, any given iteration $k$ requires $n-k$ column-wise
operations, and each of these operations are independent from
each-other. Therefore, the Householder matrix calculations can be done
in a thread-safe manner.

The second addition we made to the standard SRIF is the inclusion of a
secondary $\chi^2$ minimization for the scaling of $\vec{\delta{x}}$,
so that \[Q' = Q(\vec{x} + \alpha \vec{\delta x})\] is minimized over
$\alpha$.  This minimization is done with an eleven point grid search
for $\alpha$, from $\alpha=10^{-3} $ to $\alpha=10^{3.5} $.  The
additional minimization adds a trivial additional computation cost to
the overall minimization of $\chi^2$, but allows for faster
convergence, and the possibility of skipping over local minima in the
$\chi^2$-space.

The final change we made to the underlying SRIF algorithm also granted
the largest speed improvement.  Even with the optimizations described
above, the O$(n^2(m-n))$ nature of the triangularization algorithm
scales the computational cost drastically with increased model
complexity. Furthermore, the need to store a derivative matrix for
each iteration results in sizeable memory overhead when working with
large datasets.  To mitigate this problem, we modified the SRIF
algorithm to select a subset of the nominal free parameters during
each parameter vector adjustment, and to only fit for that subset. We
tried a variety of subset selection methods, and concluded after
testing that a ``semi-random" mode was the most effective.  During
each parameter vector adjustment, this mode randomly selects a fixed
number, $b$, of parameters from the nominal set of parameters
$\{x_s\}$ for which the condition
\[ k_s \leq \lfloor{i * \frac{b}{n}}\rfloor\]
is satisfied, where $k_s$ is the total number of times parameter $x_s$
has been considered over the course of the entire fit, $i$ is the
total number of times that the parameter vector has been adjusted, $n$
is the total number of nominal free parameters, and $\lfloor \
\rfloor$ is the round down operator.

When fitting for both shape and spin state simultaneously
(Section~\ref{sec-spin}), the spin axis orientation parameters were
always included in the fit at each parameter vector adjustment step.

\subsection{Penalty functions}
The SPF routine can currently fit models to data while taking into
account a suite of ``penalty functions" that favor models with
desirable properties.  In a way, these penalty functions serve to make
the fit operate in a more global context -- there may be a local
minimum in $\chi^2$-space towards which the fitting algorithm would
want to tend, but that minimum can be ruled out \textit{a priori}
thanks to physical considerations.  These penalty functions include
limits on ellipsoid axis ratios to avoid absurdly elongated or
flattened shapes, constraints on shape concavities to avoid
unrealistic surface topographies, and limits on the model center of
mass distance from the image center, to name a few. We have
implemented these penalty functions in the SRIF
framework
by redefining the residual vector as
\[ \vec{R}'' = \left( \begin{array}{c}  \vec{z} - \vec{f}(\vec{x}) \\  \vec{p_w} \end{array} \right),\]
where
\[ \vec{p_w} = \left( \begin{array}{c} 
p_1\times w_1 \\ \vdots \\ p_N\times w_N \end{array} \right)\] for
which $\{p_i\}$ , $\{w_i\}$ are the set of penalty functions and
penalty weights, respectively, and
\[A'' = \frac{\partial \vec{R''}}{\partial \vec{x}}.\]
The algorithm then progresses as described in section~\ref{sec-srif},
with $\vec{R}''$ replacing $\vec{R}$ and $A''$ replacing $A'$.

\section{Results}
\label{sec-results}

We tested our implementation with three different types of
data. First, we generated simple spherical harmonic shapes and
simulated images with Gaussian noise.  Second, we used existing shape
models of asteroids and simulated images with $\chi^2$-distributed
errors, the appropriate model for radar noise.  Third, we used an
actual asteroid radar data set.  In all cases, we fit the images to
recover the shapes using SRIF, SPF, and a third-party
Levenberg-Marquardt algorithm (LM), a standard optimizing algorithm
which is used across a wide variety of fields and applications
\citep{numericalrecipes}.  
Except where otherwise noted, our tests did not involve adjustments to
parameters controlling the radar scattering law, ephemeris corrections, or
spin axis orientation.

\subsection{Simulated data with artificial shapes}
\label{sec-fake}

These tests consisted of generating an initial basic shape (either
spherical, oblate ellipsoid, or prolate ellipsoid), and randomly
perturbing the spherical harmonic representation of this shape to get
a new, non-trivial object. 

\begin{figure*}[t!]
\begin{center}
a)
\includegraphics[width=0.7\columnwidth]{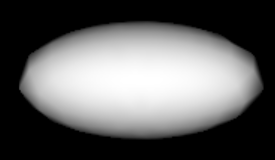}
b)
\includegraphics[width=0.63\columnwidth]{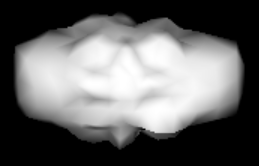}
\caption{Example of artifical shape used as a test object to be fitted
  for.  a) The initial shape, a prolate ellipsoid, before any of the
  spherical harmonic parameters have been changed. b) A perturbed
  prolate ellipsoid.
}
\label{fig-beforeafter}
\end{center}

\end{figure*}

Simulated range-Doppler images of this object were generated, and
these images were fit for using the three aforementioned algorithms.
This test serves as a good absolute test of a fitting method, because
a solution is guaranteed to exist within the framework used (namely, a
spherical harmonic representation). Figure~\ref{fig-beforeafter} shows
an example of the resulting shape when starting with a prolate
ellipsoid. Three randomly generated objects were created for each of
the three basic shapes, for a total of nine test cases.

For each test case, we then generated between 20 and 30 simulated
radar images and added Gaussian noise such that pixel values on the
target exceed the root-mean-square (rms) deviations of the noise by an
average factor of $\sim 5$ and a peak factor of $\sim 150$.  These
images were used to attempt to reconstruct the perturbed shape, with
the original basic shape (sphere, prolate ellipsoid, or oblate
ellipsoid) given as the initial condition.  This process was repeated
for each of the nine test cases.

The three fitting algorithms shared 
the same starting conditions for each test. For each fit, the models
comprised 121 free parameters (corresponding to the coefficients of a
ten-degree spherical harmonic representation), and the simulated
images contained a total of 2.4 million data points. 
Stopping criteria were also normalized for the three different test
types -- a fit was considered finished if the $\chi^2$ statistic had
not changed to within three significant digits after one hour, or
twelve hours had passed since the fit began, whichever occurred
first. Time-based stopping criteria -- as opposed to iteration-based
-- were chosen in order to account for fundamental differences between
the algorithms with respect to the definition of a single
iteration. In addition, fits were allowed to run past the criteria
stopping point, and the criteria were analyzed and applied
afterwards. This was to avoid missing a drop-off in $\chi^2$ in one
test type that might not appear in another. All times are wall clock
time.

The results of these tests (figure~\ref{fig-ninetests}) indicate that
SRIF consistently performs better than the currently-used SPF
algorithm. In addition, SRIF appears to ultimately converge on a lower
chi-squared than LM in all cases.

SRIF also converged on a reduced chi-squared ($\chi_{red}^2$) of less than 1.3
(indicating a reasonable approximation of the correct model parameters
had been found) in eight out of the nine tests, while SPF was only
able to do so in one out of the nine tests.

\subsection{Simulated data with existing asteroid shape models}
\label{sec-real}

We conducted another set of tests using existing shape models of
asteroids.  Three cases were tested - Itokawa, the 1999 KW4 primary,
and 2000 ET70 (figure~\ref{fig-sky}). As opposed to the previous set
of tests, these shapes are not guaranteed to be well approximated with
a spherical harmonic representation.  However, a best-fit spherical
harmonic representation can still be found.

For this set of images we used $\chi^2$-distributed errors, which is
the correct noise model for individual images of radar echo power.  We
chose a noise model such that the pixel values on the target exceed
the rms deviations of the noise by an average factor of $\sim 2$ and a
peak factor of $\sim 60$.  When multiple images are summed, one can
often rely on the fact that errors approach normality by the central
limit theorem, hence the choice of Gaussian noise for the images
tested in section~\ref{sec-fake}.

The results of these tests are illustrated in
figure~\ref{fig-chierrs}, 
and an example comparison of a fit synthetic image to the simulated
image 
is shown in figure~\ref{fig-obsvsfit}.  Our implementation of SRIF
clearly performs faster and with higher accuracy than both SPF and LM.

\subsection{Spin state}
\label{sec-spin}

Jointly solving for spin state and shape is typically challenging and
time-consuming with the traditional implementation of {\tt shape}.  A
common approach is to estimate the spin state as best as possible with
rudimentary shapes (e.g., ellipsoids or low-order spherical harmonic
models) in a basic grid search.  One can then use the most favorable
trial values of the spin state to fit a model for the physical shape
to the observed radar images.  Experience with traditional {\tt shape}
indicates that the algorithm rarely deviates much from the initial
conditions given on the spin state, probably as a result of the
one-parameter-at-a-time fitting approach.

Our tests indicate that SRIF is capable of fitting a reasonable
asteroid shape, even when the initial shape and spin state parameters
are far away from their optimal values.  This advantage likely results
from the joint estimation of shape and spin parameters.

For example, figure~\ref{fig-spin} shows the best-fit spherical
harmonics shape, as determined by SRIF, for a set of simulated images
of the asteroid Itokawa. The initial conditions for the shape
parameters were a sphere with a radius $10\%$ larger than the longest
axis of the actual shape model. In addition, the initial spin axis was
$30$ degrees off from the spin axis with which the data were
simulated. We repeated these experiments with a variety of starting
conditions, as well as several different shape models, with similar results.

SRIF's capacity to fit for both shape and spin state parameters can
drastically cut down on the total time required to obtain an accurate
asteroid shape model.

\subsection{Real data: 2000 ET70}

We have run {\tt shape} with all three fitting algorithms on actual
radar images of the asteroid 2000 ET70 \citep{naidu2013}.  {\tt shape}
was run initially with an ellipsoid model. The starting conditions for
this model were such that the ellipsoid axes were all equal.
The best fit ellipsoid model 
($a/b=1.16, b/c=1.13$) was then converted into a spherical harmonic
model with 122 model components (corresponding to the coefficients of
a ten-degree spherical harmonic representation, as well as one overall
size scaling factor), which was then fit again to the data. This
process resulted in a final spherical harmonic model
(figure~\ref{fig-et70fit}) for the asteroid with a $\chi_{red}^2$ of
2.1.  The stopping criterion was a reduction in $\chi_{red}^2$ less
than 0.01 between two consecutive iterations.

For the first stage, SRIF fit a substantially better ellipsoid than
SPF did, although it took about eight minutes longer
(Table~\ref{tab-etstats}).  For the second stage, SRIF converged on a
final solution more than two times faster than SPF.  This further
corroborates the results obtained from our tests with simulated data.

\begin{table*}
\centering
\caption{Run statistics for SPF and SRIF fits for 2000 ET70 data.}
\begin{tabular}{ |l|r|r|r|r|r|r|r| }
\hline \hline
Model & $\chi^2_{initial}$ & \multicolumn{3}{c|}{$\chi_{final}^2$} & \multicolumn{3}{c|}{Runtime (hours)}\\ [0.5ex]
\cline {3-8}
{} & {} & SPF & LM & SRIF & SPF & LM & SRIF \\ [0.01ex]
\hline
2000 ET70: Ellipsoid  & 3.7 & 2.8 & 2.7 & 2.4 & 0.02 & 0.84 &  0.129 \\
2000 ET70: Sph. Harm. & 2.4 & 2.10 & 2.37 & 2.10 & 2.98 & 0.1725 & 1.36\\
\hline
\end{tabular}
\label{tab-etstats}
\end{table*}

\section{Future changes}
\label{sec-future}
The addition of SRIF to {\tt shape} has improved fitting performance,
but additional changes can still be made to allow {\tt shape} to
function optimally with real-world data.

\subsection{Global vs local variable partitioning}  

The fits discussed in this paper were performed on global parameters
only -- namely, parameters that are valid across all data sets
associated with the object in question. When performing a
high-fidelity fit on multiple data sets, however, it is necessary to
take into account local parameters as well. These are model arguments
which apply only at specific points in time. For example, while the
mean radius and rotation rate of an asteroid is a global parameter,
the system temperature and ephemeris correction parameters on the
third day of observations are local to the data taken on the third day
of observations.
 
Processing local parameters is less computationally intensive than
processing global parameters.  The gradients of any observables not
within a local parameter's timeframe are known \textit{a priori} to be
zero. This greatly reduces the number of modelling function calls that
must be made when considering local parameters. In addition, the
triangularization of a derivative matrix scales with the number of
non-zero elements. This means that while the total number of
additional parameters scales like the product of the number of
datasets with the average number of local parameters per dataset, the
additional computation time only scales with the average number of
local parameters per data set. Because of this, adding the capacity
for processing local parameters will only increase runtime by
${\sim}20\%$. We plan on adding this functionality in a future version
of {\tt shape}.

\subsection{Additional fitting methods}
Tests that we have run with the simulated and real data have indicated
that the $\chi^2$-space for shape-models is not smooth.
Figure~\ref{fig-chispace} shows a two-dimensional slice of the
$\chi^2$-space for a spherical harmonic model against 2000 ET70
data. The multi-valleyed nature of this space makes it difficult for
local fitting methods to find the global minimum. In light of this,
global fitting mechanisms such as simulated annealing or Markov Chain
Monte Carlo may be better suited for this problem.  These methods can
be supplemented by a gradient descent method like SRIF. In fact,
utilizing a hybrid of these methods may prove to be the optimal
solution for this class of problem. Until such methods are
implemented, convergence on a global minimum will be dependent on a
good choice of starting conditions.
This often forces shape modelers to explore a variety of initial
conditions, and identifying such starting conditions is not always practical.

\subsection{Additional shape representations}
There are serious drawbacks to using spherical harmonics to represent
the radius of an object at each latitude-longitude grid point.  Many
asteroid shapes are poorly approximated by this representation (e.g.,
the 1999 KW4 primary) and there are entire classes of shapes (e.g., banana) that can not be
described at all in this fashion.  Traditionally, this problem has
been solved by the use of vertex models, but these shape
representations typically involve a large number of parameters (i.e.,
the coordinates of three vertices per facet). We are currently looking
into new representation methods, some of which may allow for a greater
range of shapes, while at the same time cutting down on the number of
free parameters.

\section{Conclusions}
We have added new optimization procedures into the asteroid shape
modeling software {\tt shape}, enabling the use of a
Levenberg-Marquardt algorithm or a Square Root Information Filter.  We
implemented several optimizations to the SRIF algorithm to increase
performance in shape inversion problems.  Tests on both simulated and
actual data indicate that our additions allow shape inversion to
proceed more quickly and with better fidelity than was previously
possible.  
The SRIF implementation also facilitates simultaneous fits of the spin
axis orientation and shape.

\newpage
\appendix
\section{Numerical Stability}
\label{sec-app}
Issues with numerical stability can arise when multiplying matrices
with elements at or near the square root of the machine precision. 
This can lead to erroneous results, or singular matrices for which further
calculations (such as the matrix inverse) are impossible. 

For example \citep{bierman1977}, consider the case when

 \[ A = \left( \begin{array}{c c}
1 & 1 - \epsilon \\
1 - \epsilon & 1 \\
1 & 1 \end{array} \right) \]
and
\[W = I.\]
Then  
\[ (A^\intercal W A) = \left( \begin{array}{c c}
3 - 2\epsilon + \epsilon^2 & 3 - 2\epsilon \\
3 - 2\epsilon & 3 - 2\epsilon + \epsilon^2 \end{array} \right) 
 \]
 Thus, in the case that $\epsilon$ is equal to or
less than the square root of the machine precision, 
 \[(A^\intercal W A) = \left( \begin{array}{c c}
3 - 2\epsilon & 3 - 2\epsilon \\ 3 - 2\epsilon & 3 - 2\epsilon
\end{array} \right).\] This matrix is
singular, and thus $(A^\intercal W A)^{-1}$ cannot be computed. This
problem is particularly insidious because matrix singularity in higher
dimensions can be difficult to detect. 

\clearpage

\begin{figure*}
\begin{center}

\includegraphics[width=0.9\columnwidth]{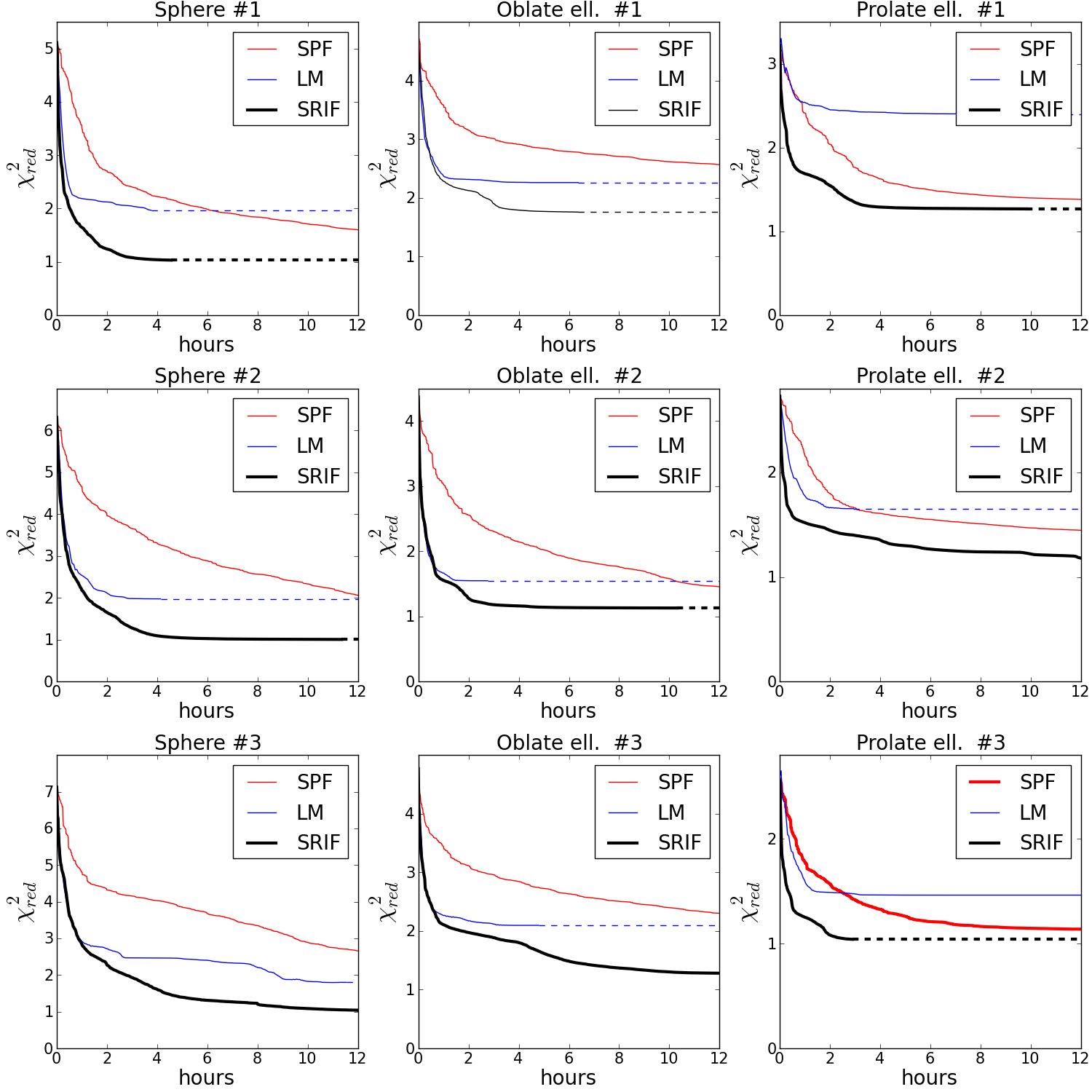}
\caption{Results of three fitting algorithms (Sequential Parameter
  Fit, Levenberg-Marquardt, and Square-Root Information Filter) with
  three artificial shapes (perturbed versions of a sphere, oblate
  ellipsoid, and prolate ellipsoid).  Bold lines indicate fits which
  converged to a $\chi^2_{red} < 1.3$. Dashed lines indicate the
  assumed future state for fits that had converged on a solution
  before the 12-hour time frame.}
\label{fig-ninetests}
\end{center}
\end{figure*}

\clearpage

\begin{figure*}
\begin{center}
a) \includegraphics[width=0.27\columnwidth]{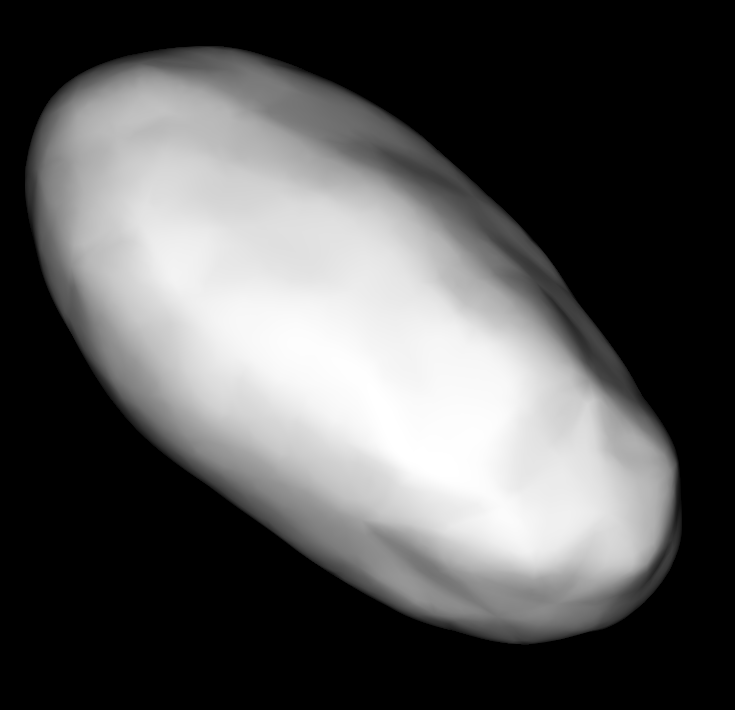}
b) \includegraphics[width=0.3\columnwidth]{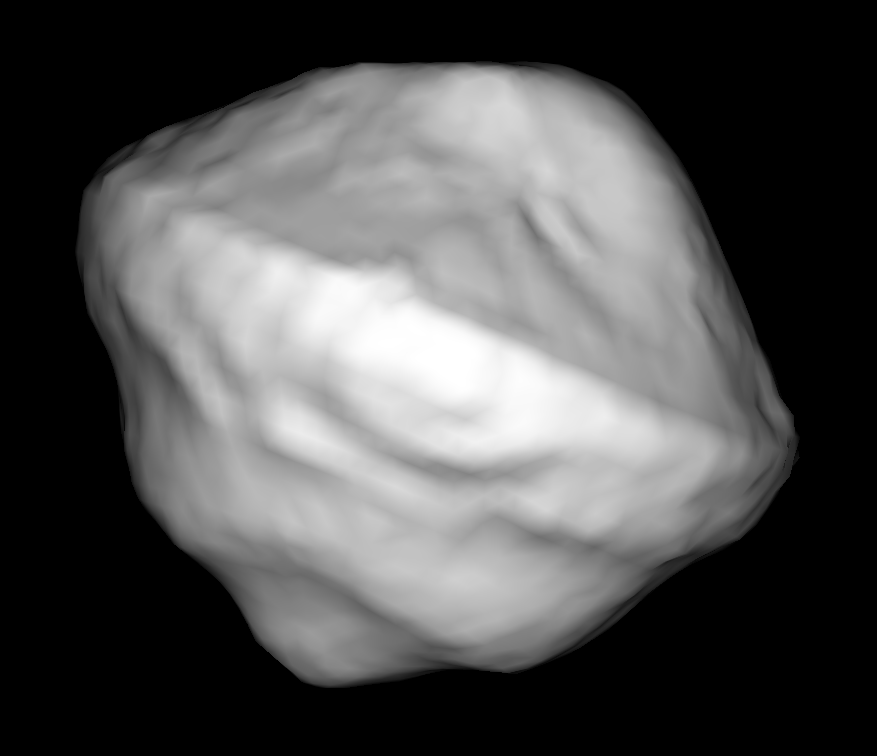}
c) \includegraphics[width=0.31\columnwidth]{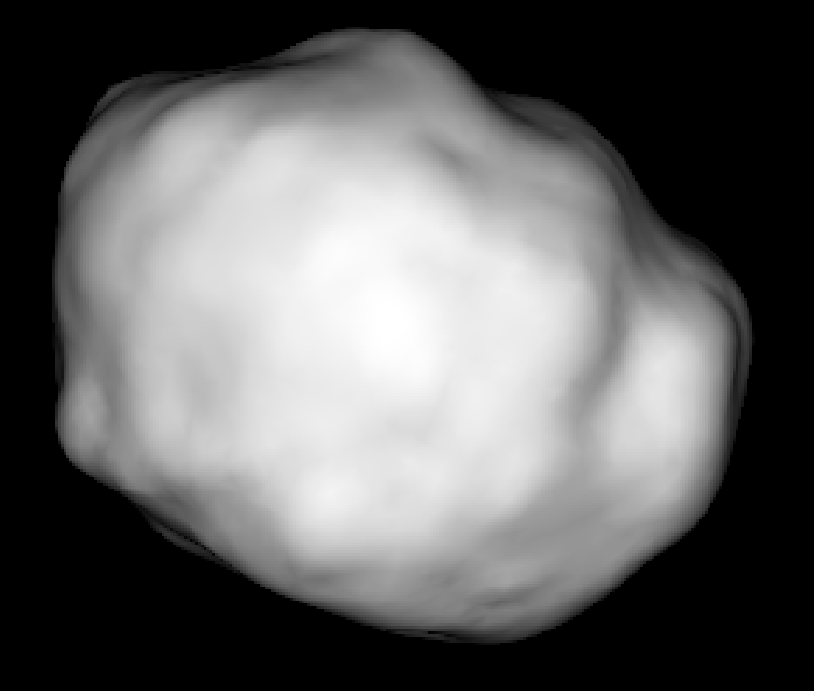}
\caption{Plane of sky representations of the radar-derived shape models of (a) Itokawa, (b) 1999 KW4, and (c) 2000 ET70 .}
\label{fig-sky}
\end{center}
\end{figure*}

\begin{figure*}
\begin{center}
\includegraphics[width=1.0\columnwidth]{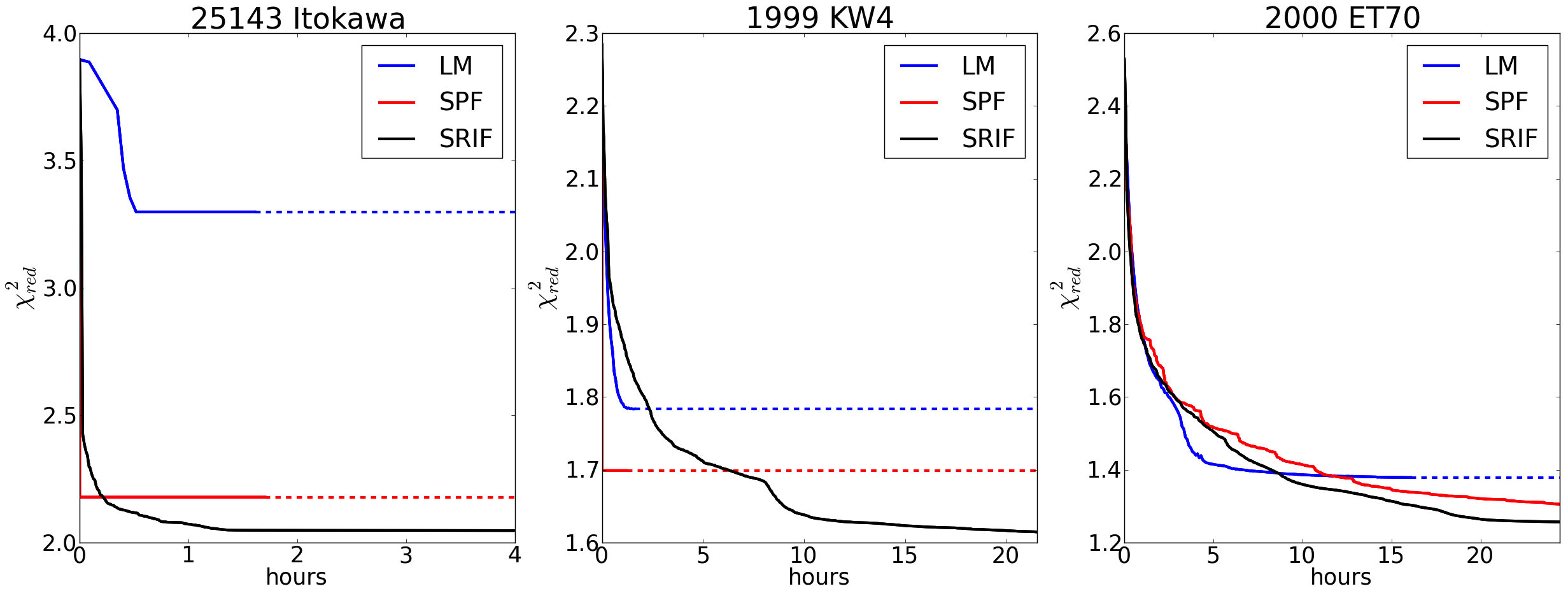}
\caption{Results comparing SRIF, LM, and SPF operating on real asteroid
  shapes with simulated $\chi^2$-distributed errors.  Dashed lines
  indicate the assumed future state for fits that had converged on a
  solution. These fits were run without penalty functions.
Note that the solution arrived at by SPF for 1999 KW4 was a
non-physical, pebble-sized asteroid. 
Avoiding non-physical minima in the $\chi^2$ space would require human
intervention to manually tweak the starting conditions.  We did not
perform such tweaks in order to maintain consistency in our tests.}
\label{fig-chierrs}
\end{center}
\end{figure*}

\begin{figure*}
\begin{center}
a) \includegraphics[width=0.42\columnwidth]{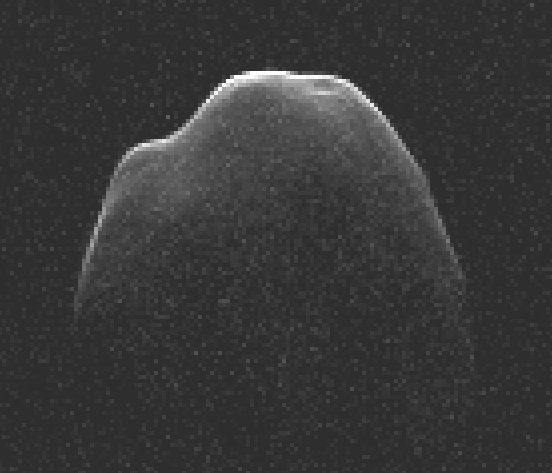}
b) \includegraphics[width=0.42\columnwidth]{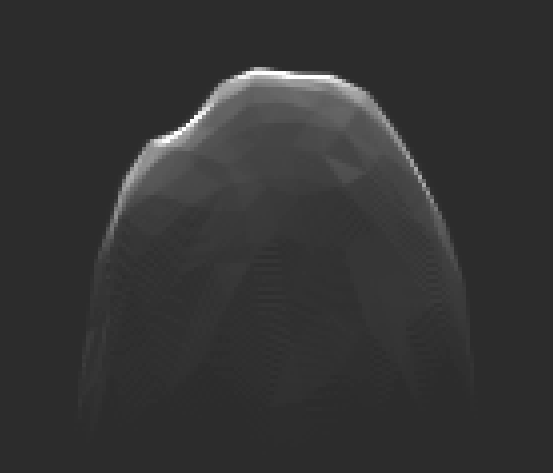}
\caption{(a) Example of a simulated input to the shape modeling
algorithm.  The input is generated by projecting the shape model into
range-Doppler space at a specific observation epoch and adding random
noise. (b) The corresponding synthetic image produced by the shape
modeling algorithm after fitting for the shape.}

\label{fig-obsvsfit}
\end{center}
\end{figure*}

\begin{figure*}
a) \includegraphics[width=0.95\columnwidth]{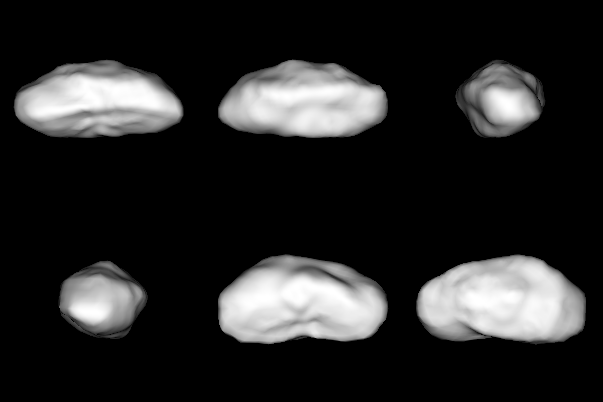}

b) \includegraphics[width=0.95\columnwidth]{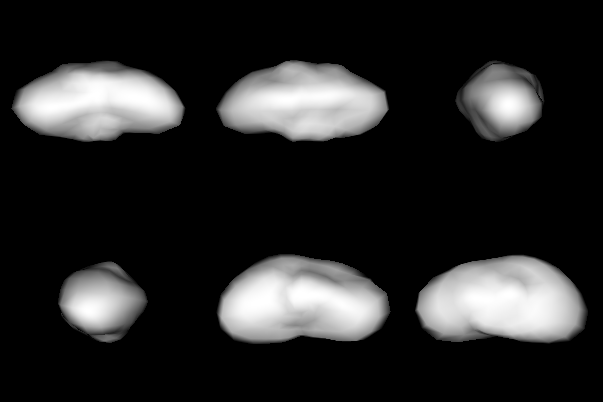}
\caption{ a) The Itokawa shape model that was used to generate
  simulated radar images \citep{ostro05}.
  b) The best-fit SRIF
  tenth degree spherical-harmonics model for those simulated data, using penalty functions.
  The initial
  conditions for the shape parameters were a sphere with an offset spin axis 
  latitude and longitude. }
\label{fig-spin}
\end{figure*}

\begin{figure*}[h!]
\begin{center}
\includegraphics[width=0.8\columnwidth]{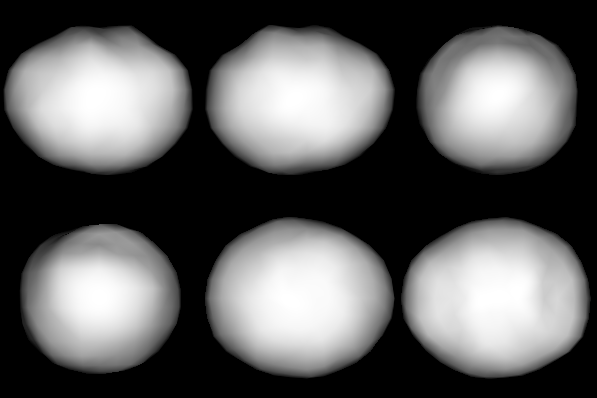}
\caption{
The 
best SRIF fit spherical harmonic model for 2000 ET70, which
is in good agreement with the model of \citet{naidu2013}.  }
\label{fig-et70fit}
\end{center}
\end{figure*}

\begin{figure*}
\begin{center}
\includegraphics[width=0.8\columnwidth]{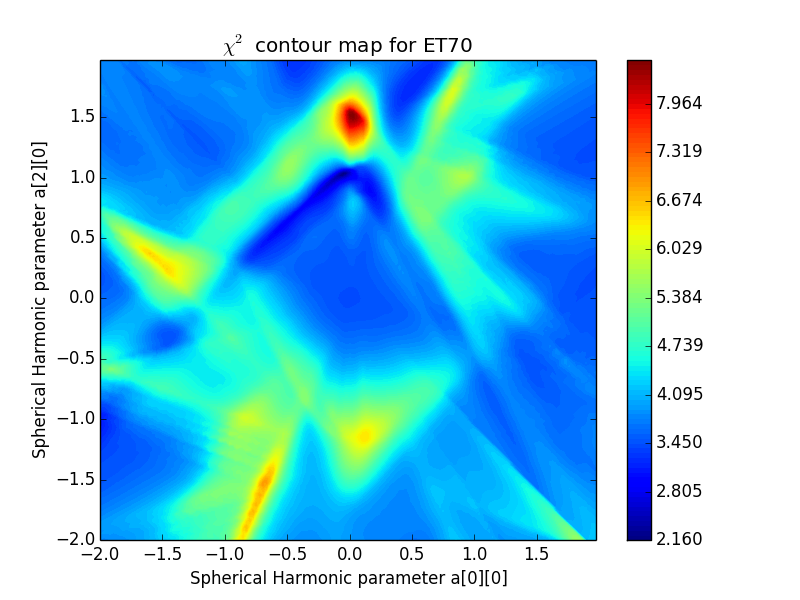}
\caption{A two-dimensional slice of the $\chi^2$ space for fitting
  shape models to radar data. This contour map represents $\chi^2$ for
  a spherical harmonic model with all parameters fixed except two of
  the elements of the primary coefficient matrix.  Note that the blue
  regions indicate low $\chi^2$, and that there are several of these
  regions for a derivative-based optimizer to fall towards, depending
  on the initial set of starting conditions.}
\label{fig-chispace}
\end{center}
\end{figure*}

\FloatBarrier

\clearpage
\bibliography{comps_bib}
\end{document}